# Tailoring Kinetics on a Topological Insulator Surface by Defect-Induced Strain: Pb Mobility on Bi$_2$Te$_3$

**Wen-Kai Huang,**[†] **Kai-Wen Zhang,**[†] **Chao-Long Yang,**[†] **Haifeng Ding,**[†,‡] **Xiangang Wan,**[†,‡] **Shao-Chun Li,**[†,‡*]

[†]National Laboratory of Solid State Microstructures, School of Physics, Nanjing University, Nanjing 210093, P. R. China, [‡]Collaborative Innovation Center of Advanced Microstructures, Nanjing University, Nanjing 210093, P. R. China

**James W. Evans,**[§,‖] **and Yong Han**[§,‖,*]

[§]Department of Physics and Astronomy, Iowa State University, Ames, Iowa 50011, United States, [‖]Ames Laboratory—U. S. Department of Energy, Iowa State University, Ames, Iowa 50011, United States

*Address correspondence to
S.-C.L.
Email: scli@nju.edu.cn; Tel: +86-(0)25-83596050
Y.H.
Email:  yong@ameslab.gov; Tel: 1-515-294-4819

## ABSTRACT

Heteroepitaxial structures based on Bi$_2$Te$_3$-type topological insulators (TIs) exhibit exotic quantum phenomena. For optimal characterization of these phenomena, it is desirable to control the interface structure during film growth on such TIs. In this process, adatom mobility is a key factor. We demonstrate that Pb mobility on the Bi$_2$Te$_3$(111) surface can be modified by the engineering local strain, $\varepsilon$, which is induced around the point-like defects intrinsically forming in the Bi$_2$Te$_3$(111) thin film grown on a Si(111)-7×7 substrate. Scanning tunneling microscopy observations of Pb adatom and cluster distributions and first-principles density functional theory (DFT) analyses of the adsorption energy and diffusion barrier $E_d$ of Pb adatom on Bi$_2$Te$_3$(111) surface show a significant influence of $\varepsilon$. Surprisingly, $E_d$ reveals a cusp-like dependence on $\varepsilon$ due to a bifurcation in the position of the stable adsorption site at the critical tensile strain $\varepsilon_c \approx 0.8\%$. This constitutes a very different strain-dependence of diffusivity from all previous studies focusing on conventional metal or semiconductor surfaces. Kinetic Monte Carlo simulations of Pb deposition, diffusion, and irreversible aggregation incorporating the DFT results reveal adatom and cluster distributions compatible with our experimental observations.

The discovery of topological insulators (TIs) has drawn intense attention due to the intriguing properties of this new type of quantum matter.[1-3] Owing to the helical spin-momentum texture of topological surface state protected by time reversal invariance, and the insulating bulk band structure, TIs find promising potential in many fields, such as quantum computing[4] and spintronics[5]. Electronic structure of TIs has recently been extensively studied.[6-11] Many exotic topological phenomena have been revealed at interfaces with a TI , e.g., the emergence of Majorana fermions at a TI-superconductor interface,[12-14] the quantum anomalous Hall effect in a magnetic TI,[15-16] and the prediction of charge triggered image magnetic monopoles near a magnetic layer at a TI surface,[17] etc. In order to achieve an in-depth understanding of those novel phenomena arising at TI-related surfaces or interfaces, well-controlled structure, particularly at atomic scale, is vitally important.[13, 18-22] Very recently, experimental observations have revealed that grain boundaries in $Bi_2Se_3$ thin film growing on SiC(0001) can cause surface strain, and corresponding density functional theory (DFT) calculations show that this strain can significantly influence the topological Dirac states,[23,24] DFT calculations for strain effects on energy band structures for bulk or thin film forms of TIs have been also reported.[25-28] However, engineering of kinetics at TI surface, e.g., tuning surface mobility using strain, has yet to be reported. The most critical factor in controlling film morphology is adatom surface mobility, which is exactly the focus of this study. High mobility produces further separated larger islands, thus reducing the density of domain boundaries formed upon coalescence. Higher mobility also facilitates interlayer transport and thus film smoothness. Both features can have a significant impact on properties of TI surface or metal-TI interface.

In this study, we demonstrate experimentally for the first time the modification of adsorption and diffusion of metal adatoms on a TI surface by exploiting strain effects. Specifically, we deposit Pb on $Bi_2Te_3$, a typical TI. In addition, the study of this system is motivated in part by intensive efforts searching for Majorana Fermions which may exist at a superconducting Pb-TI interface.[29] By using scanning tunneling microscopy (STM), we show that point-like defects can intrinsically form at the surface of a thin $Bi_2Te_3(111)$ film grown on a Si(111)-7×7 substrate, and a local strain field is induced around these defects during the film growth. Adsorption, diffusion, and aggregation of deposited Pb are markedly affected by the surface strain field: isolated Pb adatom densities on both unstrained and compressed regions are much larger than that on regions with tensile strain, while larger Pb clusters preferentially form on the defects in the tensile-strained regions. To explain this behavior, we perform first-principles DFT calculations for the adsorption energies and diffusion barriers of Pb adatoms on the $Bi_2Te_3(111)$ surface with variable strain, and kinetic Monte Carlo (KMC) simulations based on the DFT results.

The influence of strain on adatom mobility has been assessed previously for surfaces of conventional metals and semiconductors. One such class of STM studies explores the dependence of adatom mobility on film thickness in lattice-mismatched heteroepitaxy.[30-32] Alternatively, uniaxial strain can be directly applied to substrates to assess effects on surface structure and kinetics.[33,34] On the theory side, many studies have explored the dependence of the adatom adsorption energy, $E_{ad}$, and surface diffusion barrier, $E_d$, on strain, $\varepsilon$, which is conventionally defined as $\varepsilon = [(a_s - a)/a] \times 100\%$, where $a_s$ ($a$) is the lattice constant of strained (unstrained) surface. $\varepsilon > 0$ means tensile strain, and $\varepsilon < 0$ compressive strain. In these studies, $E_d$ typically varies linearly and increases monotonically with increasing $\varepsilon$ over the explored range,[35-38] although distinct behavior can occasionally occur on some surfaces.[39,40]

Our DFT analysis for Pb on $Bi_2Te_3(111)$ shows that $E_d$ is not a monotonic function of $\varepsilon$ in the small-$\varepsilon$ regime and has a cusp-like minimum at $\varepsilon_c \approx +0.8\%$, in contrast to the typical behavior on conventional metal or semiconductor surfaces, where the $\varepsilon$-dependent $E_d$ is linear for $\varepsilon = -1\%$ to $+1\%$. This unusual behavior is shown to reflect a bifurcation or discontinuous jump in the location of the most stable adsorption site as $\varepsilon$ passes through $\varepsilon_c$. KMC simulations of Pb deposition, diffusion, and aggregation incorporating the DFT value of $E_d$, clarify the kinetic (versus thermodynamic) origin of the observed difference in adatom and cluster density on strained versus unstrained domains. The adatom and cluster distributions from KMC simulations are comparable with our experimental observations.

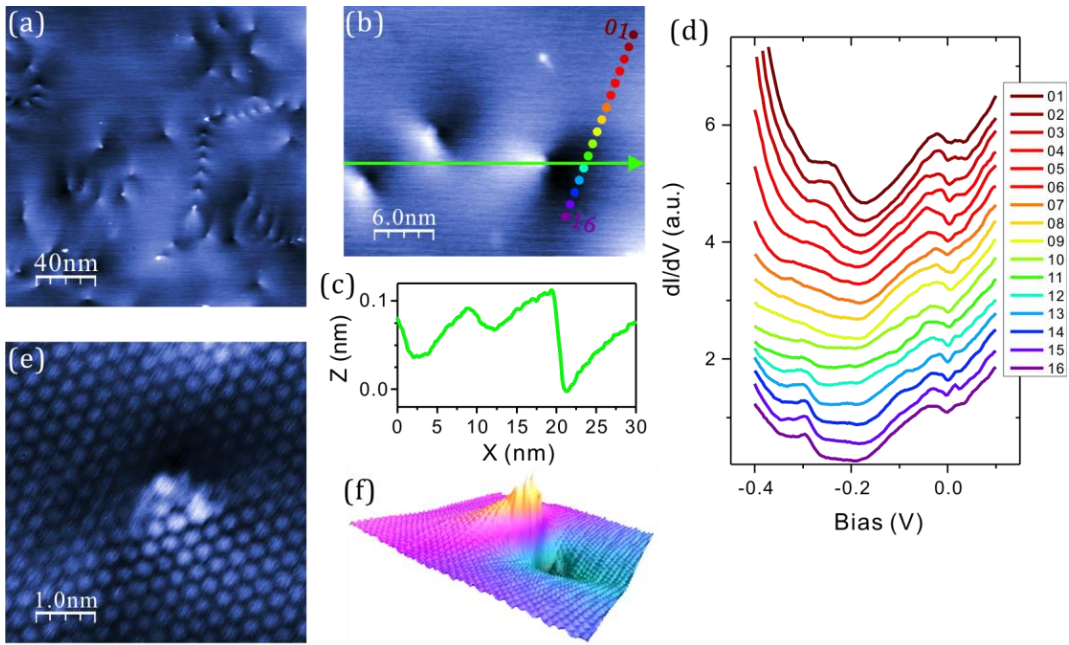

**Figure 1. (a) 200×200 nm² STM image of $Bi_2Te_3(111)$ film obtained at 80 K (tunneling current $I =$ 0.1 nA, bias voltage $U =$ +1.0 V). $Z$ scale is 0 - 0.35 nm. (b) 30×25 nm² STM image of $Bi_2Te_3(111)$ film obtained at 4 K ($I =$ 0.1 nA, $U =$ +1.0 V). (c) Line-scan profile measured along the green horizontal line in (b). (d) d$I$/d$V$ spectra taken at 4 K along the rainbow-colored dotted line in (b), sweeping from unstrained region to strained region. Colored dots from 01 to 16 in (b) correspond to the same colored curves in (d), respectively. (e) A typical point-like defect with atomic resolution (image size: 5×5 nm²; $I =$ 0.2 nA, $U =$ +5.0 mV). (f) 3D view of surface corrugation around a point-like defect. Image size: 20×20 nm².**

Figure 1a shows the surface of a thin (~10 nm) $Bi_2Te_3(111)$ film grown on a Si(111)-7×7 substrate. In contrast to a perfect flat surface reported for much thicker $Bi_2Te_3$ films,[10, 24, 41] this surface exhibits well-defined defect structures. Most of the defects are point-like, some of them occasionally lining up. A curved surface usually develops around a defect structure, as seen in Figure 1b,c. The upper surface of top quintuple layer (QL, see Figure S1 in Supporting Information) of the film is bent up and down around the defect, and thus develops stretched zones. The corrugation of this bent surface is in total ~0.07 nm. Simple atomistic or continuum level analysis of strain, as well as previous experimental studies,[42] indicate that the protruding regions reflect lateral tensile strain. Detailed examination of an atomic-resolution STM image (Figure 1e and Figure S2) shows how the atoms are laterally displaced and stretched relative to a perfect substrate lattice. The stretching is most prominent close to the defect center, and decays gradually away from the center. Typical lattice strain is roughly estimated to be less than ~1.0% beyond ~5 nm from the defect center (see Figure S2). The vertical strain is not detectable. Our geometric phase analysis (GPA)[23] indicates a complicated strain field close to the defect center: tensile and compressive regions are alternatively distributed. However, on a larger length scale (from 3 - 20 nm away from the defect center), the strain is mainly tensile and gradually decays, see Figure S4. This behavior is similar to observations in other systems.[42] The defect-induced surface strain field is also revealed in scanning tunneling spectroscopy (STS) measurements, see Figure 1d. STS measurements in the non-stretched zone (see 01 in Figure 1d) shows a clear valence band top and conductance band bottom, and a nearly linear shape in between as the hint of topological surface state, in agreement with previous reports,[10, 43-44] whereas spectroscopy in the strained region shows clearly a gap-like feature (see 16 in Figure 1d), in line with other experiments and theoretical calculations for the TIs under pressure.[23, 26-27]

The origin of these defects is not immediately clear from STM images. However, previous work by Liu et al.[23] found that low-angle tilt grain boundaries (GBs) in the top QL with tilt angles $\theta < 15°$ and $<\theta> \approx 8°$ naturally produce quasi-periodic strings of such point defects. Fast Fourier transform (FFT) analysis of our STM data around more isolated defects indicates that these may be associated with tilt GBs with even lower tilt angles $\theta \leq 3°$. Other possibilities may also exist for the structure of such isolated defects, e.g., that they are induced by subsurface GB, see Figure S3.

Our experimental observations also show that the population of point-like defects on the $Bi_2Te_3(111)$ surface can be varied by controlling growth conditions, e.g., deposition rates, film thickness, and selection of substrate. For illustrations, Figure S5 shows a series of defective surface morphologies of $Bi_2Te_3(111)$ films with different deposition rates, film thicknesses and substrates.

After depositing ~0.005 monolayer (ML) Pb on the above $Bi_2Te_3(111)$ surface at temperature $T \approx 100$ K and flux (deposition rate) $F \approx 0.005$ ML/s, Pb adsorbs predominantly as single adatoms (monomers) with occasional formation of small clusters, see Figure 2a. High-resolution imaging, as shown in Figure 2d, indicates that a Pb monomer is preferentially located at the fcc site surrounded by three surface Te atoms (see inset in Figure 3a). The slight dark halo is also observed around each protrusion of Pb monomer on $Bi_2Te_3$, which is possibly due to slight charge transfer to substrate, see Figure 2f.

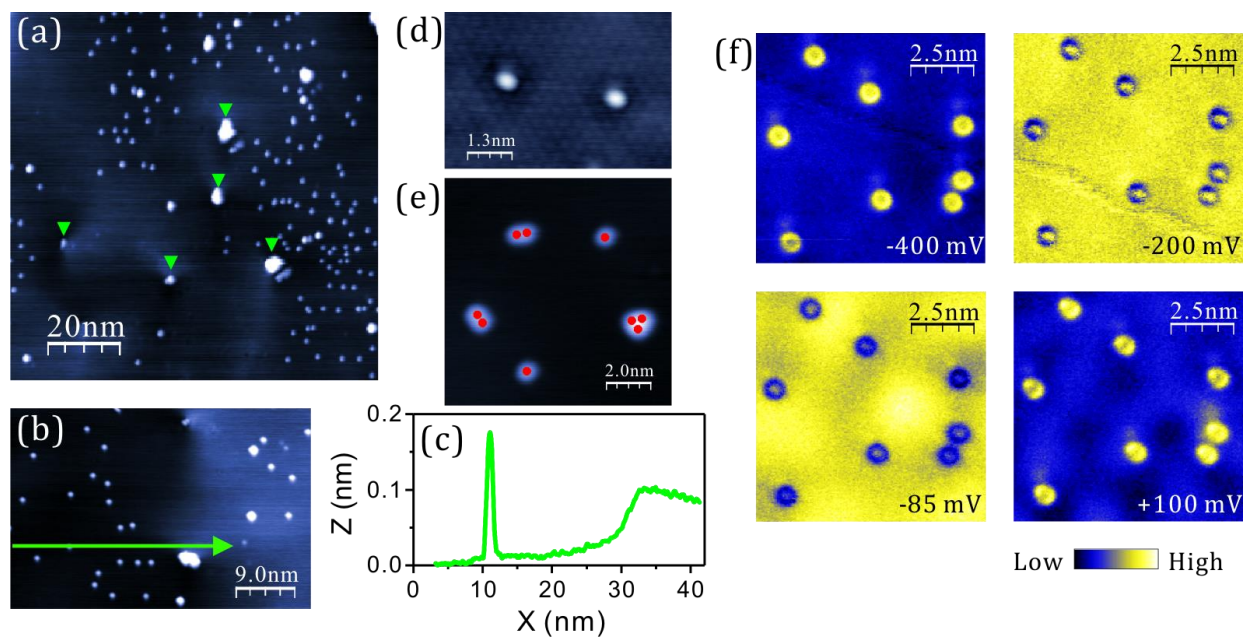

**Figure 2.** (a) 100×100 nm² STM image (*I* = 0.1 nA, *U* = +1.0 V) for depositing Pb on Bi₂Te₃(111) surface with point-like defects. *Z* scale is 0 - 0.75 nm (including the height of Pb clusters). Green arrows mark the positions of the defects. (b) 46×30 nm² STM image (*I* = 100 pA, *U* = +1.0 V) of transition from flat region (left) to bent-up tensile region (right). (c) Line-scan profile taken along the green arrow in (b). (d) 6.5×4.5 nm² atomic-resolution STM image (*I* = 220 pA, *U* = +9.0 mV) of Pb monomers on an unstrained region. (e) 10×10 nm² STM image (*I* = 100 pA, *U* = +1.0 V) identifying Pb monomers, dimers and a trimer on an unstrained region. A red dot indicates a Pb atom. The Pb coverage is ~0.005 ML for (a,b,d) and ~0.01 ML for (e). (f) STS (d*I*/d*V*) mapping taken on a few Pb monomers with various voltages. All the images are obtained at ~80 K, and STS (d*I*/d*V*) mapping at ~4 K.

On a larger scale, a rather non-uniform distribution of Pb is produced on the surface. The adsorption of Pb is highly correlated with the locations of the point-like defect structures. In general, the Pb monomers prefer to adsorb in the flat (unstrained) or depressed (compressive) zones, while their population is depleted in most of the protruding zones. Instead, small clusters form within the zones with depleted population of single Pb atoms and larger clusters sometimes form at their periphery, where these clusters have presumably captured many atoms deposited in the protruding zones. Figure 2b shows the transition region between flat and protruding zones illustrating the correlation between local surface curvature and Pb adsorption: Pb monomers populate the flat zone (left side in Figure 2b), and larger clusters are found in the strained zone (right side in Figure 2b). Figure 2c, a line scan profile measured along the green line in Figure 2b, indicates the local surface curvature and the apparent height of a Pb adatom. Note that the clusters near the depleted zones are usually larger than dimers and trimers (also see Figure S7,8). This phenomenon indicates that diffusion is enhanced in the broader neighborhood (3 - 20 nm) of the defect center where tensile strain is introduced. Furthermore, the behavior in the flat (unstrained) and tensile zones differs.

We also find that even though single Pb atoms constitute the dominant configuration at low coverage on unstrained regions, Pb dimers, trimers and even tetramers are occasionally observed as well. Figure 2e shows a mixture of monomers, dimers, and trimers observed at coverage of 0.01 ML. Naturally, the relative population of clusters (size $s \geq 2$ atoms) is boosted with increasing Pb coverage, also see Figure S8.

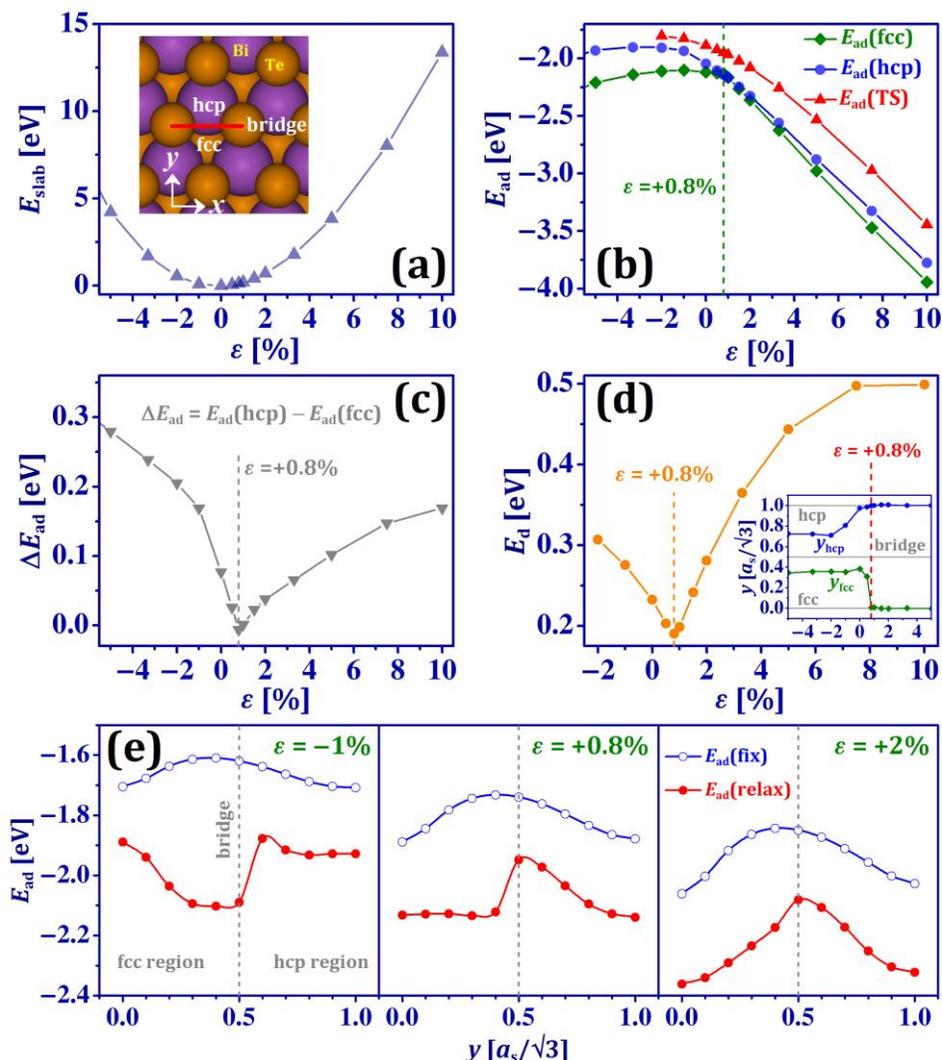

**Figure 3. DFT energetics versus strain parameter $\varepsilon$. (a)** Total energy $E_{slab}$ of 2-QL slab without Pb adatom. Inset: fcc, hcp, bridge regions on $Bi_2Te_3(111)$. The origin of $x$ and $y$ coordinates is chosen at a high-symmetry fcc site. **(b)** Pb adsorption energies $E_{ad}$ at most stable sites in the fcc and hcp regions, as well as at the TS for hopping. **(c)** Adsorption energy difference $\Delta E_{ad} = E_{ad}(hcp) - E_{ad}(fcc)$. **(d)** Diffusion barrier $E_d$. Inset: location $y$ (in units of $a_s/\sqrt{3}$, the distance between high-symmetry fcc and hcp sites) of most stable sites in the fcc and hcp regions. The origin of coordinates is chosen at the high-symmetry fcc site. **(e)** Cut through PES at $x = 0$ for varying $y$ showing seesaw variation versus $\varepsilon$ with (red solid circles) and without (blue open circles) substrate relaxation. See also Figure S10 in Supporting Information.

Our DFT analysis for surface diffusion of Pb on $Bi_2Te_3(111)$ aims to elucidate the above behavior. The strained surface is modeled by changing the surface lattice parameter $a$ to $a_s$ where the bulk $a$ is determined from DFT analysis, see Supporting Information. Figure 3a shows the curve of the total energy $E_{slab}$ of a clean 2-QL 3×3 slab versus $\varepsilon$. The minimum of $E_{slab}$ occurs at $\varepsilon = 0$, i.e., the unstrained surface.

We first calculate the adsorption energies $E_{ad}(fcc)$ and $E_{ad}(hcp)$ of a Pb adatom at the lowest-energy location in the fcc and hcp regions, which are shown in the inset of Figure 3a. Here $E_{ad} = E_{tot} - E_{slab} - E_{Pb}$, where $E_{tot}$ is the total energy of slab with the Pb adatom, and $E_{Pb}$ is the energy of one Pb atom in gas phase. Negative values of $E_{ad}$ reflect the attractive adatom-slab interaction. From Figure 3b, $E_{ad}(hcp) = E_{ad}(fcc)$ at $\varepsilon \approx +0.8\%$, while both $E_{ad}(fcc)$ and $E_{ad}(hcp)$ are quasi-linearly decreasing (i.e., the adsorption becomes stronger) with increasing $\varepsilon > +0.8\%$. However, behavior for $\varepsilon < +0.8\%$ is strongly non-linear. Figure 3c highlights this non-linearity showing $\Delta E_{ad} = E_{ad}(hcp) - E_{ad}(fcc)$. Here it should be also mentioned that, after relaxation of a Pb adatom in the fcc region on unstained $Bi_2Te_3(111)$, the optimized height of the Pb adatom relative to the top Te atom center is ~0.19 nm, which is in good agreement with the experimental value of ~0.17 nm, as indicated by the line-scan profile for the apparent height for single Pb atom in Figure 2c.

Next, to obtain $E_d$ values, using configurations obtained above for relaxed Pb in fcc and hcp regions, we perform climbing nudged elastic band (cNEB) calculations to obtain the minimum energy paths for Pb motion between these sites, and thus the diffusion barrier $E_d$, which is defined as the difference between the energy at the transition state (TS) for hopping and $E_{ad}$ for the most stable adsorption site. Results for $E_d$ versus $\varepsilon$ are shown in Figure 3d. Surprisingly, $E_d$ has a cusp-like minimum of ~0.19 eV at $\varepsilon = \varepsilon_c \approx +0.8\%$. Such non-monotonic behavior for $\varepsilon = -1\%$ to 1% is quite different from that in metal or semiconductor systems reported previously.[35-40] We track the unusual non-analytic cusp-like behavior in both $\Delta E_{ad}$ and $E_d$ versus $\varepsilon$ to a bifurcation in the location of the stable adsorption site in the fcc region. For $\varepsilon > \varepsilon_c$ (tensile strain), this stable site is close to the high-symmetry fcc site. However, as $\varepsilon$ drops below $\varepsilon_c$ (including compression), it jumps to a location closer to the bridge site. See Figure 3d inset. This behavior is in turn tracked to a seesaw-like bifurcation in the quasi-linear potential energy surface (PES) for adsorption in the fcc region which tilts down towards the fcc site for $\varepsilon > \varepsilon_c$, but towards the bridge site for $\varepsilon < \varepsilon_c$. See Figure 3e. A more complete set of PES profiles capturing this behavior is shown in Figure S10, and also see more details in Supporting Information. Such novel PES behavior is more likely on complex binary alloy surfaces.

The above seesaw-like bifurcation behavior is associated with the complexity of the substrate TI surface, a binary alloy with hexagonal close-packed intralayer structure, where the top, second, and third single atomic layers alternate between Te, Bi, and Te, respectively. We expect this type of behavior is possible more generally for binary alloy surfaces (not specifically for TI), and will depend on the relative strength of binding of the adatom to the constituents, and also on geometric aspects of the system, as well as on substrate atom relaxation. The dramatic change in behavior of the PES with and without relaxation of substrate atoms shows the importance of this effect.

Figure 4 shows the DFT preferred adsorption sites of Pb adatom on $Bi_2Te_3(111)$ surface for three different $\varepsilon$. The shift in the stable adsorption site for the Pb adatom (cf. Figure 3d inset) has already been discussed. To illustrate the magnitude of the relaxation of surface Te atoms, we take a Te atom marked "S" in Figure 4 as an example by checking the change $\Delta y$ (in unit of $a_s/\sqrt{3}$) of its position along the $y$ direction before and after full relaxation. For $\varepsilon = -2\%$, $\Delta y = 0.228$; for $\varepsilon = 0$, $\Delta y = 0.281$; for $\varepsilon = +2\%$, $\Delta y = 0.000$. This indicates that the relaxation of the surface Te atom is large for both compressive strain $\varepsilon = -2\%$ and no strain ($\varepsilon = 0$), while there is no relaxation for the large tensile strain $\varepsilon = +2\%$. In addition, we also perform a Bader charge analysis[45] for three configurations in Figure 4. We find that there is indeed a charge transfer of about $0.5 - 0.6e$ from the Pb adatom to the $Bi_2Te_3(111)$ substrate within a 3×3 lateral unit cell, and the charge distribution is rather localized around the Pb adatom. This DFT result is consistent with the above charge-transfer expectation from the experimental observations in Figure 2d,f.

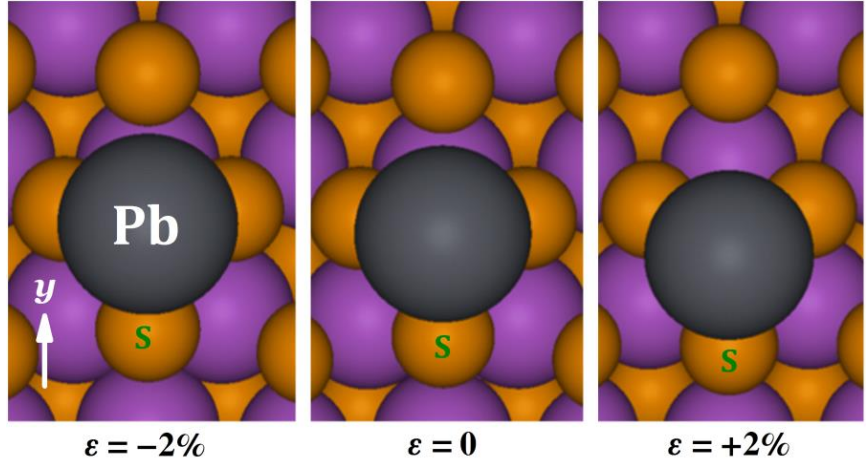

**Figure 4. Top views for DFT preferred adsorption sites of Pb adatom on $Bi_2Te_3(111)$ surface for three different $\varepsilon$ after full relaxation.**

As a consequence of the above analysis, we conclude that the higher density of Pb monomers and small clusters on the unstrained region is due to the corresponding higher $E_d$. The lower population of Pb in the protruding regions with tensile strain (which we estimate to be around 0.8%) must reflect the lower $E_d$. Beyond this qualitative picture, we have performed KMC simulations[46-47] of Pb deposition, diffusion, and irreversible aggregation into clusters for a range of $h/F$ values. We assume the diffusion is thermally activated and the hop rate $h = \nu e^{-E_d/(k_B T)}$ with typical value $\nu = 10^{12}$/s, where $k_B$ is the Boltzmann constant. We also allow post-deposition evolution for up to a few minutes to mimic the experimental protocol. Results for the density of monomers, dimers, trimers, and all clusters ($s \geq 2$ atoms) versus $h/F$ after deposition of 0.01 ML Pb are shown in Figure 5a,b. Choosing the DFT value $E_d \approx 0.23$ eV for the unstrained surface corresponds to $h/F \approx 4 \times 10^2$ at a deposition temperature $T \approx 100$ K. The KMC results for unstrained surface are qualitatively comparable to the experimentally observed populations of monomers, dimers, and trimers, as shown in Figure 5c,d. On the other hand, for the strained region, $E_d \approx 0.19$ eV is lower and $h/F \approx 5 \times 10^4$ is higher corresponding to the negligible population of Pb monomers.

As an aside, to assess whether Pb adatom clustering on the $Bi_2Te_3(111)$ surface is reversible or not, we calculate the pair interactions $E_{ff}$ ($E_{hh}$) for two Pb adatoms on nearest-neighbor fcc (hcp) sites. Our DFT results from a $3 \times 3$ cell are $E_{ff} = -0.429$ eV and $E_{hh} = -0.545$ eV for $\varepsilon = 0$; $E_{ff} = -0.490$ eV and $E_{hh} = -0.470$ eV for $\varepsilon = +0.8\%$. Here, negative interaction values mean attraction. These attractions are strong enough to ensure that Pb adatom clustering is irreversible.

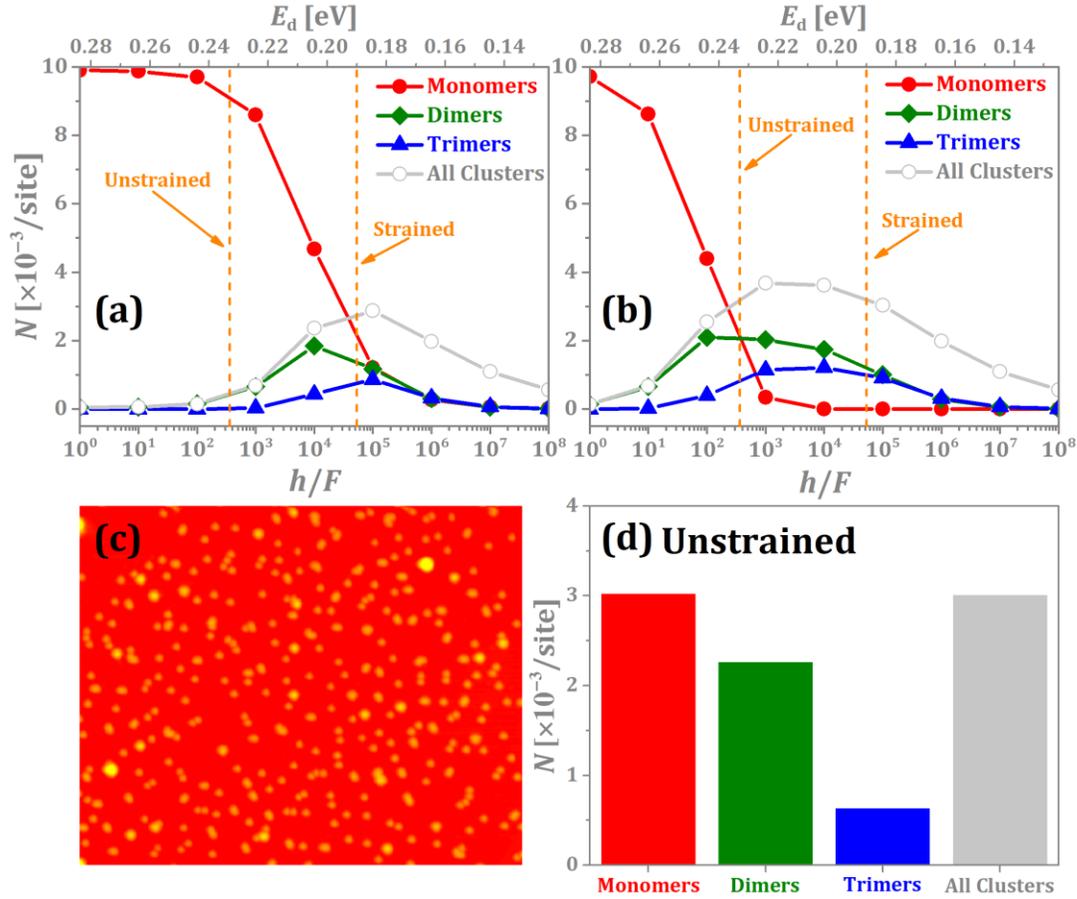

**Figure 5.** KMC simulation and experimental results for densities of monomers, dimers, trimers, and all clusters versus $h/F$ and $E_d$. Simulated densities (a) just after deposition of 0.01 ML Pb, and (b) a few minutes later. Two vertical dashed lines correspond to $E_d = 0.233$ and 0.190 eV for unstrained ($\varepsilon = 0$) and strained ($\varepsilon = +0.8\%$) Bi$_2$Te$_3$(111) surface, respectively. (c) 110×90 nm$^2$ STM image of ~0.01 ML Pb on unstrained Bi$_2$Te$_3$(111). (d) Experimental densities of Pb clusters on unstrained Bi$_2$Te$_3$(111).

In summary, to our knowledge, this work reveals, for the first time, modification of adsorption and diffusion behavior on a TI surface by manipulation of local surface strain. The population of point-like defects inducing strained regions on thin TI surface can be controllable via tuning growth conditions. Our analysis and understanding of this unexpected strain-dependent behavior provides more general insight into manipulation of atom mobility at TI surfaces and interfacial structures. This provides new opportunities to manipulate heteroepitaxial growth, and thus the interfacial geometries, which will be helpful in the search for novel quantum transition behavior at the TI-based interfaces. In addition, the wrinkled Bi$_2$Te$_3$ surfaces appearing in the thin film regime also provide a special platform for atomic level investigation of pressure-induced quantum transitions, such as topological superconductivity or gap opening of Dirac fermion surface states.

## METHODS

**Sample Preparation.** All the experiments were carried out in ultrahigh vacuum instrument equipped with a molecular beam epitaxy chamber. The base pressure is $1 \times 10^{-11}$ mbar. $Bi_2Te_3(111)$ film was grown at rate $F \approx 0.01$ ML/min on a Si(111)-7×7 substrate pre-cleaned by cycles of flash annealing up to 1000°C. Prior to growth, Si(111) substrate was outgassed overnight at 400°C and then flashing up to 1000°C for about 50 cycles. The 7×7 periodicity was monitored by *in situ* reflection of high-energy electron diffraction (RHEED), and the surface cleanliness and roughness was checked by STM. Si(111)-7×7 substrate was kept at 250°C during $Bi_2Te_3$ growth. A buffer layer of ~2 ML of Te was first deposited at Si(111). The film growth rate was set to be about 0.5 ML/min. The ratio of $Te_2$ to Bi was optimized to ~20:1.[41]

**STM Characterization.** STM characterization was performed with a Unisoku LT-STM 1600 at 80 K and 4K in constant current mode. When doing Spectroscopy measurement, a lock-in amplifier was used, a modulation signal of 10 mV and 999 Hz was applied. A mechanically polished Pt-Ir tip was used.

**DFT Calculation**. Total energy DFT calculations used the plane-wave VASP code with the PAW method and PBE-GGA functional. Surface diffusion involves hopping between fcc and hcp site regions with diffusion barrier $E_d$ obtained from the cNEB method. For more details, see Supporting Information.

*Conflict of Interest:* The authors declare no competing financial interest.

*Acknowledgment.* The work at Nanjing University is supported by the State Key Program for Basic Research of China (Grants No. 2014CB921103, No. 2013CB922103) and National Natural Science Foundation of China (Grants No. 11374140, No. 11374145). Y.H. and J.W.E. are supported by NSF grants CHE-1111500 and CHE-1507223 utilizing NERSC, XSEDE, and OLCF resources.

*Supporting Information Available:* Crystal structure of $Bi_2Te_3$ (Figure S1), lattice constants of bulk $Bi_2Te_3$ obtained from different functionals in DFT analysis and from experiments (Table S1), a point-like defect in greater detail on the $Bi_2Te_3(111)$ film (Figure S2), surface and subsurface domain boundaries of the thin $Bi_2Te_3(111)$ film grown on Si(111)-7×7 substrate (Figure S3), FFT and GPA strain field maps for a strained surface area (Figure S4), various defective morphologies of $Bi_2Te_3(111)$ film obtained by controlling growth kinetics (Figure S5), STSs for strained and unstrained surface regions (Figure S6), additional images of the surface morphology after depositing a small amount of Pb (Figure S7,8), sequential STM images for hopping of a Pb adatom (Figure S9), and a more complete set of PES profiles (Figure S10). This material is available free of charge via the Internet at http://pubs.acs.org.